# The Operationalization of "Fields" as WoS Subject Categories (WCs) in Evaluative Bibliometrics: The cases of "Library and Information Science" and "Science & Technology Studies"

Loet Leydesdorff [*a] & Lutz Bornmann [b]

[*] Corresponding author
[a] Amsterdam School of Communication Research (ASCoR), University of Amsterdam, Kloveniersburgwal 48, 1012 CX Amsterdam, The Netherlands; loet@leydesdorff.net .

[b] Division for Science and Innovation Studies, Administrative Headquarters of the Max Planck Society, Hofgartenstr. 8, 80539 Munich, Germany; bornmann@gv.mpg.de .

**Abstract**

Normalization of citation scores using reference sets based on Web-of-Science Subject Categories (WCs) has become an established ("best") practice in evaluative bibliometrics. For example, the Times Higher Education World University Rankings are, among other things, based on this operationalization. However, WCs were developed decades ago for the purpose of information retrieval and evolved incrementally with the database; the classification is machine-based and partially manually corrected. Using the WC "information science & library science" and the WCs attributed to journals in the field of "science and technology studies," we show that WCs do not provide sufficient analytical clarity to carry bibliometric normalization in evaluation practices because of "indexer effects." Can the compliance with "best practices" be replaced with an ambition to develop "best possible practices"? New research questions can then be envisaged.

**Introduction**

The Subject Categories of the Web of Science (WoS) have increasingly evolved from a classification scheme for the retrieval into a standard for normalizations in bibliometric evaluations. Because publication and citation practices can be expected to differ among fields of science, one cannot compare units across fields without proper classification of "like with like" (Irvine & Martin, 1984): one should not compare apples with oranges or, in other words, books from the humanities with journal articles in fields with rapid research fronts such as biochemistry (Price, 1970). But since most units under evaluation are mixed in terms of their disciplinary composition, baselines are needed for the comparison.

The impact factor (IF) cannot be used for this evaluation because, like other bibliometric indicators, this measure varies itself systematically among fields of science. Garfield (1972) introduced the IF deliberately with a citation window of only the last two years of cited articles in order to focus on activities at the research front (Bensman, 2007; cf. Martyn & Gilchrist, 1968); but not all fields entertain research fronts to the same extent or by using similar communication channels (Leydesdorff, 2008).

Schubert, Glänzel, and Braun (1986; cf. Schubert & Braun, 1986) introduced the normalization of citation rates as relative to expected citation rates. Using a subset of 400 physics journals, these authors defined the so-called relative citation rate (RCR) as the mean observed citation rate divided by the mean expected citation rate, where the latter is based on the average in a reference set. Relative citation rates can then be computed for subsets. Normalization against the average citation rate of similar publications (in terms of document types) in the same journal provides an obvious candidate for the delineation of a reference set, but normalization at the field level requires composed sets of journal literature. The unambiguous classification of journals using citation matrices, however, has remained an unsolved problem in bibliometrics (Leydesdorff, 2006; Rafols & Leydesdorff, 2009).

Boyack & Klavans (2011) noted that many journals are not sufficiently disciplinarily organized to be used as units of analysis for the normalization. Bradford's Law of Scattering (1934) and Garfield's (1971) Law of Concentration predict that topics are scattered over journals: subject sets are inherently fuzzy and cannot be semantically defined by words (Bensman, 2001). Schubert, Glänzel, and Braun (1989, at p. 7; cf. Braun *et al*., 1994) already noted that "the field/subfield classification of papers is a neuralgic point of all kind of scientometric evaluations." These authors used the classification system of Computer Horizons Inc. (CHI) that is still current for the *Science and Engineering Indicators* of the National Science Board of the USA (National Science Board, 2014). Moed *et al*. (1995, at p. 386) proposed using the "ISI journal categories" as reference sets. These journal categories were mainly computer-generated on the basis of title words from the very start of the *Science Citation Index* in the 1960s.

Using the ISI journal categories for the delineation of reference sets in terms of journals, Moed *et al.* (1995) developed the measure CPP/Fcsm—citations per publication compared to the mean citation score of a field—as an addition to CPP/Jcsm, that is, the equivalent but then normalized at the level of a journal. This CPP/Fcsm was advocated by the Leiden Center for Science and Technology Studies (CWTS) as the "crown indicator." More recently, CPP/Fcsm has been replaced by a "new crown indicator": MNCS, or the "mean normalized citation score." In the Leiden Ranking 2014,[1] MNCS is most recently no longer defined with reference to journals or sets of journals, but based on categories that are algorithmically generated from citation relations among papers (Ruiz-Castillo & Waltman, in preparation; Waltman & van Eck, 2012).[2]

The ISI journal categories were renamed into the WoS Subject Categories (WC) with the introduction of the current version 5 of the Web-of-Science (WoS) in 2009 (Leydesdorff, Carley, and Rafols, 2013). Unlike the CHI-NSF classification—that is currently maintained by Patent Board™ under a contract with the NSF—more than a single WC can be attributed to each journal. In bibliometric evaluation, a journal is commonly attributed a percentage proportional to the categories under which it is subsumed. These multiple categories have also been considered as an indication of the interdisciplinarity of journals, and the overlaps among categories accordingly are assumed to exhibit the complexity ("interdisciplinarity") of the journal structures (Bordons *et al.*, 2004; Katz & Hicks, 1995; Morillo *et al.*, 2001). However, different categories may cover rather similar sets of journals; for example, in the biomedical domain (Rafols & Leydesdorff, 2009, p. 1830). In other cases, the categories added by an indexer may generate relations among otherwise unrelated journals. This can be useful for purposes of information retrieval, but blurs the analytical distinctions.

---

[1] The Leiden Rankings 2013 were still based on using WCs for the normalization.
[2] MNCS also corrects a problem in the statistics of the old indicator (Gingras & Larivière, 2011; Opthof & Leydesdorff, 2010; Waltman *et al.*, 2011). In addition to MNCS, CWTS has introduced MNCS-2 that corrects for the (deviant) first year of the citation window.

In the meantime, the use of these journal categories has become accepted as "best practice" among bibliometric practitioners (e.g., Rehn *et al*., 2014). The Flemish ECOOM unit for evaluation in Leuven, however, uses a different classification system for journals (SOOI) specifically developed by this unit (Glänzel & Schubert, 2003). Other authors have refined the journal lists within specific WCs for a more precise evaluation of a given discipline (e.g., Van Leeuwen and Calero-Medina, 2012). In the meantime, another journal classification system in terms of fields and subfields has been made available by Scopus.[3] However, we focus here on the WCs because these are so widely used for the normalization in bibliometric practices.[4]

For example, InCites—a customized, web-based research evaluation tool developed by Thomson Reuters— routinely provides normalizations of citation impact using these WCs for the delineation of the reference sets (e.g., Bornmann and Marx, 2014, at p. 496; see also Costas *et al*., 2010, at p. 1567). Since 2003, the *Journal Citation Reports* (JCR) provide also the medians of impact factors for each journal category. Using the normalization in terms of WCs, for example,

---

[3] The field/subfield classification of Scopus is available in the journal list from http://www.elsevier.com/online-tools/scopus/content-overview .
[4] Web-of-Science Subject Categories are available (under subscription) at http://images.webofknowledge.com/WOKRS56B5/help/WOS/hp_subject_category_terms_tasca.html .

Leydesdorff *et al*. (2014) and Bornmann *et al*. (in press) studied nations in terms of their contributions to the top-1% most-highly-cited publications.

Note that this delineation of reference sets in terms of journals may be pragmatic, but reference sets can also be defined in terms of (combinations of) keywords or thesauri. For example, *Chemical Abstracts* contains high-quality classification terms at the level of each paper (Bornmann *et al*., 2009; Neuhaus & Daniel, 2009), and Medline/PubMed provides a system of Medical Subject Headings (MeSH) at the paper level (Leydesdorff & Opthof, 2013; Rotolo & Leydesdorff, in press). The advantage of WoS (and Scopus), however, remains their "multidisciplinarity" in the sense that all disciplines are covered. When properly normalized, a comparison among different institutional units can thus be envisaged.

However, Garfield himself warned that the ISI—currently Thomson Reuters—assigns journals to categories by "subjective, heuristic methods" (Pudovkin and Garfield, 2002, at p. 1113n):

> …This method is "heuristic" in that the categories have been developed by manual methods started over 40 years ago. Once the categories were established, new journals were assigned one at a time. Each decision was based upon a visual examination of all relevant citation data. As categories grew, subdivisions were established. Among other tools used to make individual journal assignments, the Hayne-Coulson algorithm is used. The algorithm has never been published. It treats any designated group of journals as one macrojournal and produces a combined printout of cited and citing journal data.[5]

According to these authors, the categories are sufficient, but they added that "in many areas of research these classifications are crude and do not permit the user to quickly learn which journals are most closely related" (p. 1113). Boyack *et al.* (2005) estimated that the attributions could be correct in approximately 50% of cases across the file (Boyack, *personal communication,* 14 September 2008). Leydesdorff & Rafols (2009) concluded that the ISI Subject Categories can be used for statistical purposes, but not for a detailed evaluation. In the case of interdisciplinary fields, problems of imprecise or potentially erroneous classifications can be expected (Haustein, 2012, p. 101). Let's explore this question about the quality of the WCs empirically.

**Empirical examples: LIS and STS**

As cases to illustrate our argument about the problems with using WCs for normalization, we focus on the two fields with which we are most familiar so that we are able to validate the results: library and information science (LIS) and science and technology studies (STS). Of course, these cases are specific, but, in our opinion, all fields of science are more or less specific.

---

[5] Pudovkin & Fuseler (1995, p. 228) further specified the Hayne-Coulson algorithm as follows: "The number of citations each journal receives from different specialty core journals is obtained annually by a computer routine (Hayne-Coulson) that is used to create the JCR database."

In *JCR* 2012, 85 journals were assigned to the category "information science & library science" which is abbreviated in the classification system as "NU". These 85 journals are attributed 131 WCs or on average 1.54 WC/journal. Eighteen of the other attributions are to "computer science, information systems" (ET), 11 to the "management" category (PC), and the others occur three times or less frequently. *JASIST*, for example, is additionally assigned to the category ET ("computer science, information systems"), whereas *Scientometrics* is classified as "computer science, interdisciplinary." The *Journal of Informetrics* is uniquely attributed to "information science & library science".

One can construct an aggregated journal-journal citation matrix among these 85 journals using JCR 2012. Seventy-six of these 85 journals (89%) contain references to one of the other journals in the set.[6] All journals, however, are cited in this domain except *Informacios Tarsadalom*, a Hungarian journal founded in 2001. This journal was cited only once in the entire JCR 2012.

---

[6] The non-citing journals in 2012 are: *Annual Review of Information Science and Technology, Econtent, Informacios Tarsadalom, Information Research: An International Electronic Journal, Information Technology and Libraries, Journal of Organizational and End User Computing, Libraries & the Cultural Record, Online,* and *Scientist*. There can be different reasons for this, for example when a journal no longer exists but remains part of the cited archive (e.g., *ARIST*). Other journals (e.g., the *Scientist*) do not contain references.

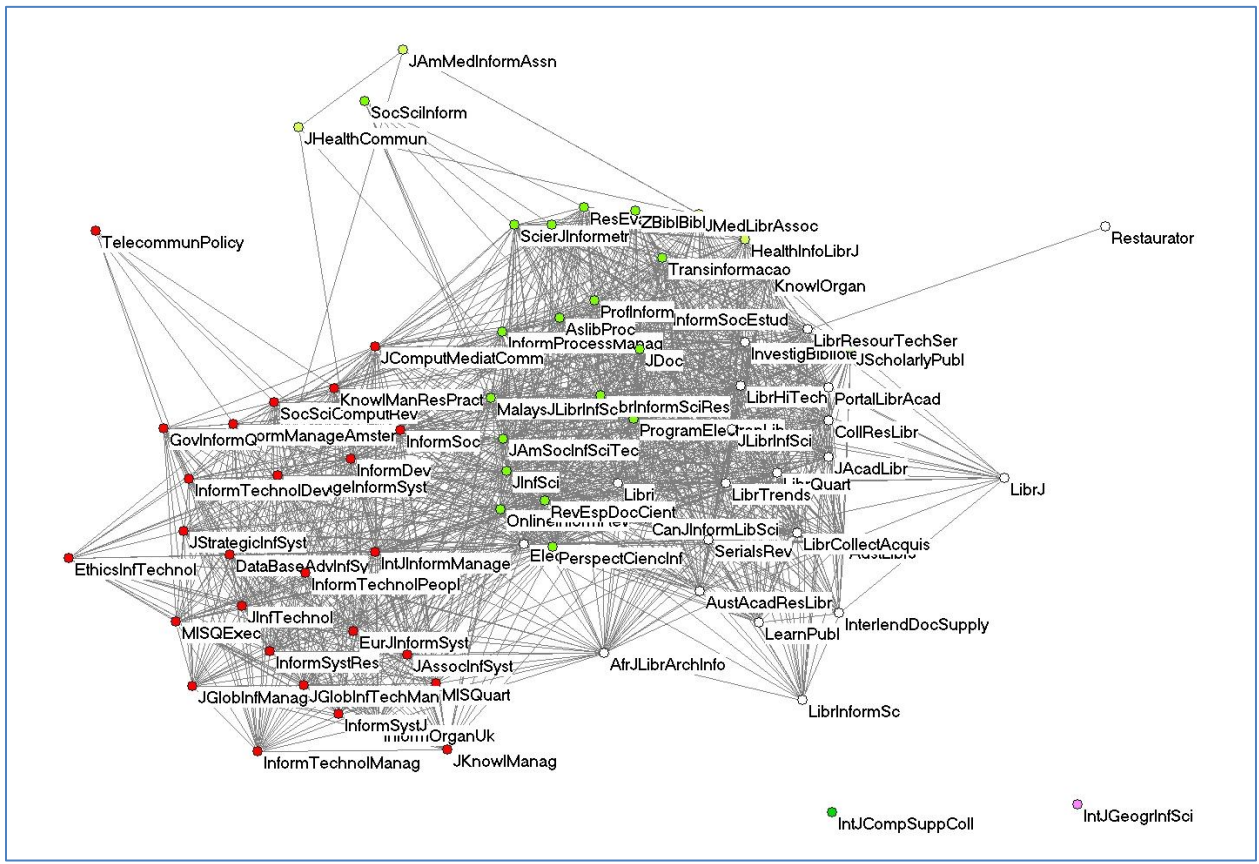

**Figure 1**: Map of 76 (of the 85) journals in the category "information science & library science" of JCR 2012 (*citing*); *cosine* > 0.05; *Q* = 0.376 (Blondel et al., 2008). Kamada & Kawai (1989) is used for the visualization.

Using a network analysis and visualization program such as Pajek or VOSviewer, one can map the journal-journal citation matrix after normalization for relative weights of the citation vectors using the cosine, and thus obtain, for example, Figure 1. Figure 1 shows how the fields and disciplinary delineations are actively reproduced by current (that is, 2012) citation behavior, whereas the cited patterns (in Figure 2) inform us about the archival structures.

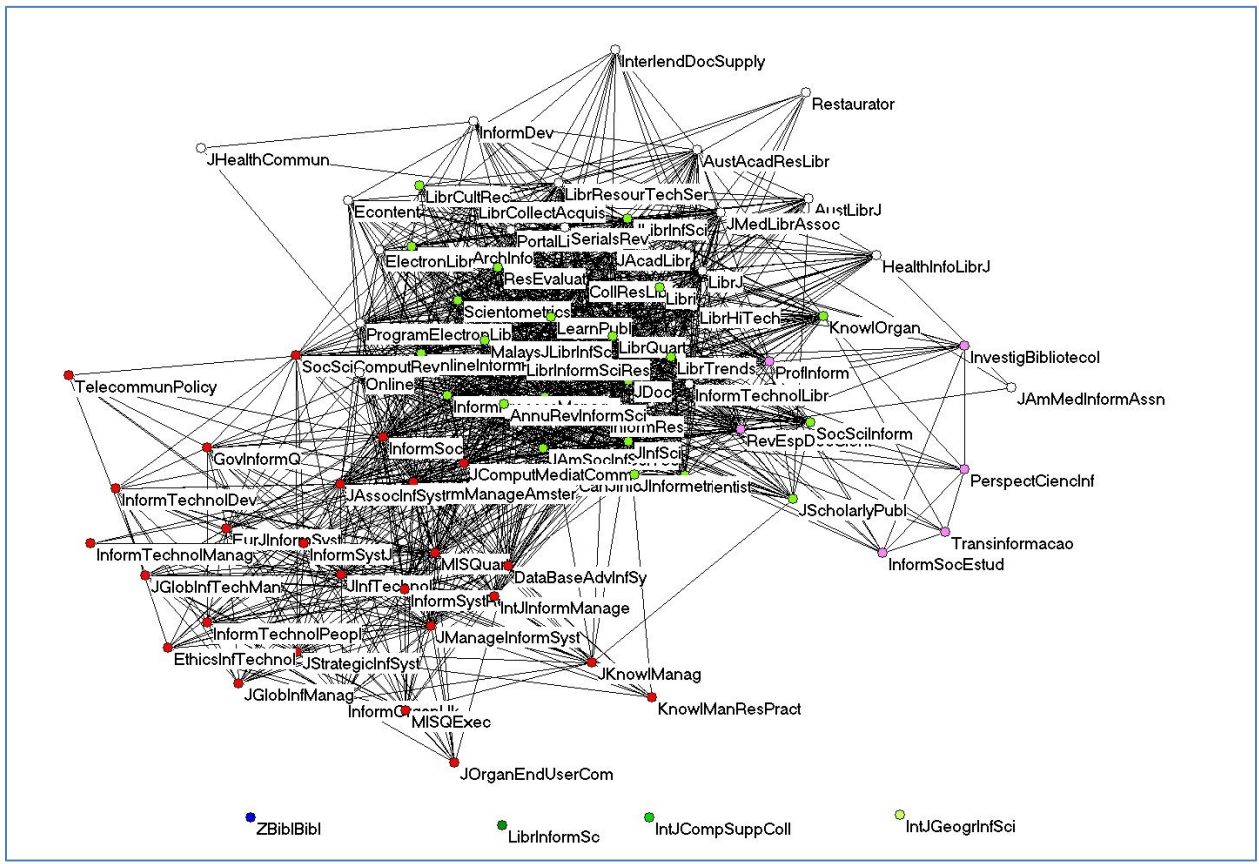

**Figure 2**: Map of 84 (of the 85) journals in the category "information science & library science" of JCR 2012 (*cited*); *cosine* > 0.05; *Q* = 0.318 (Blondel et al., 2008). Kamada & Kawai (1989) was used for the visualization.

Both maps clearly exhibit a major divide between the LIS journals, on the one side, and the information-systems journals (red-colored nodes), on the other. In the latter group *MIS Quarterly* is the leading journal with IF(2012) = 4.659, whereas *Journal of Informetrics* has the highest IF(2012) = 4.153 among the LIS journals. However, these two journals did not (!) cite each other in 2012.

One would expect the second category "ET," which stands for "computer science, information science," to be indicative of this division; but, as noted, this classification is also attributed to *JASIST* which is a central journal of LIS. Of the 84 journals in Figure 2, 27 are classified as belonging to the second partition (red-colored nodes), but only 8 of these 27 are attributed to the WC "Computer science, information systems" (among them *MIS Quarterly*). However, *Information Systems Journal, Information Systems Research*, and *MIS Quarterly Executive* are not provided with this attribution.

The LIS group itself shows a fine structure of 25 journals in the information sciences (green-colored nodes) and 22 in library science (white-colored nodes). Using another methodology, Waltman *et al*. (2011) distinguished three subsets of journals in this group: information science (14 journals), library science (27 journals), and scientometrics (9 journals). In Figure 2, six journals form a separate partition of journals in Spanish and Portuguese (pink-colored in Figure 2). These latter journals are indistinguishable from the larger set when analyzing their referencing patterns, but they are cited differently. Thus, a community-finding algorithm (Blondel *et al*., 2008) classifies them with the other LIS journals when focusing on citation behavior. In evaluative bibliometrics, however, one is interested in normalizing in terms of "being cited," and not in "citing" behavior (Nicolaisen, 2007; Wouters, 1998). Thus, the difference in citation when writing in languages other than English is relevant for the normalization.

In summary, the attribution of WCs to journals can be confusing. The classification does not work properly for the normalization even in more detailed cases. For example, when defining "informetrics" (or "iMetrics"; see Milojević and Leydesdorff, 2013) in terms of *Journal of Informetrics* (*JoI*)*, Scientometrics*, and a subset of *JASIST*, current evaluation practices would count *JoI* for 100% in the reference set because it is designated as "NU" exclusively as a single class, whereas the other two journals are each normalized for 50% with reference to this set and for 50% with reference to two other sets. Perhaps, this makes no significant differences among distributions of large units (such as countries; cf. Glänzel, 2010), but this process seems insufficiently precise for a professional evaluation at the institutional or individual level.

**Science and technology studies**

Many scholars in evaluative bibliometrics consider "science and technology studies" (STS) or more broadly "science, technology, and innovation studies" as their professional identity although methodologically scientometrics has become part also of the information sciences. The Centre for *Science and Technology Studies* (CWTS) in Leiden, for example, is hosting a major conference in this field under the title "19[th] International Conference on Science and Technology Indicators," in September 2014.

How can this professional identity be appreciated in the evaluation? Unlike LIS, STS cannot be identified by a single WC. Using citation analysis at the level of journals, STS can be defined differently from various angles (Leydesdorff & Van den Besselaar, 1997), and although institutionalized to some extent, it can also be considered as transient (Van den Besselaar, 2001; Leydesdorff, 2007) or meta-stable since one translates continuously between the interdisciplinary specialty of STS and the mother disciplines (such as the sociology of science and technology, history and philosophy of science, business and management, etc.). However, the quantitative side of STS (including scientometrics) has become increasingly important in recent decades. Using journals and citations in a set of routines for the classification of 109,164 potentially relevant articles published between 1956 and 2012, Milojević *et al*. (2014, p. 696; cf. Martin *et al*., 2012) were able to show that quantitatively oriented STS studies have in the meantime become more important in terms of numbers of publications than qualitative STS (Figure 3).

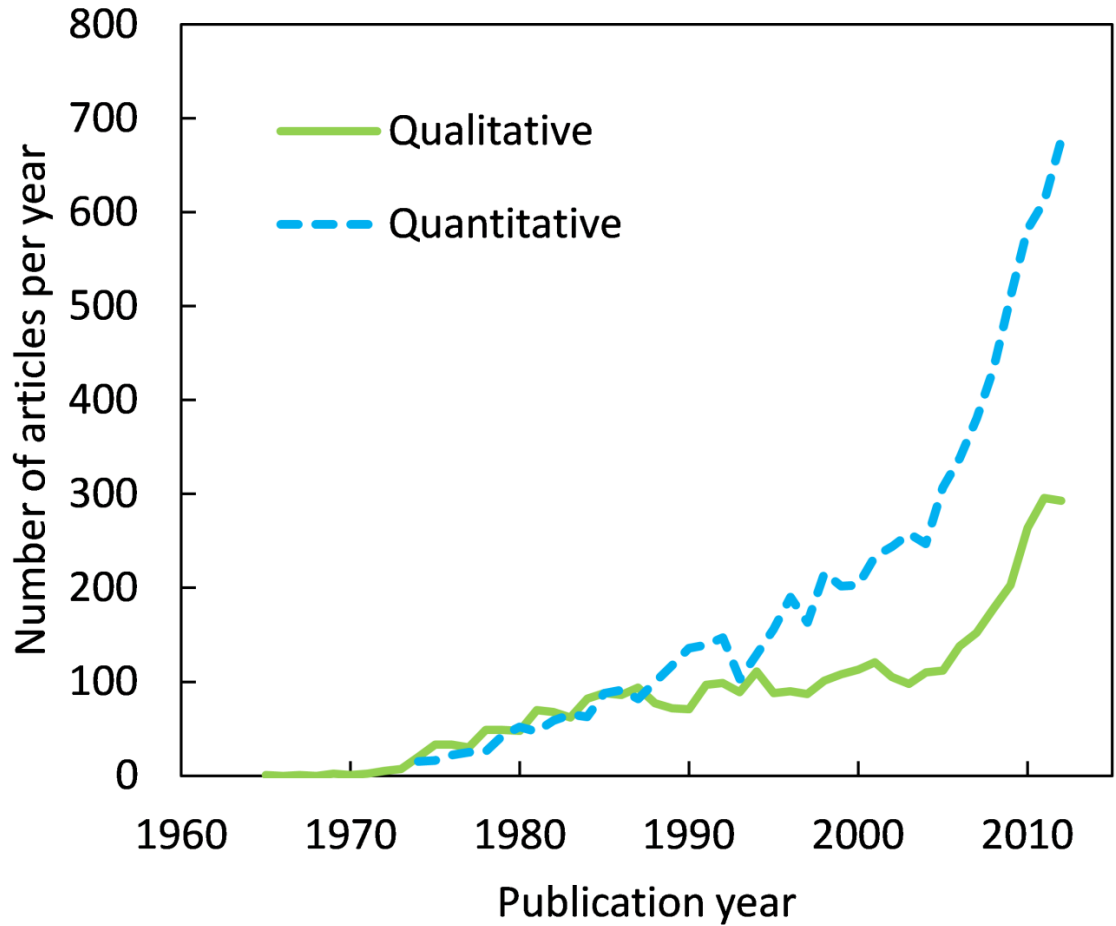

**Figure 3**: Number of journal articles in the area of STS published between 1965 and 2012 (Source: Milojević *et al.*, 2014, p. 696).

Although it may be difficult to delineate STS unambiguously in terms of journals, the WC contains a category "Social science, interdisciplinary" (WU). One would expect many of the relevant journals to be in this category. Table 1, however, tells us a different story.

|  | *WoS Category* | *Abbreviation* |
|---|---|---|
| *Scientometrics* | Information science & library science | NU |
|  | Computer sci., interdisciplinary | EV |
| *Social Studies of Science* | History and philosophy of science | WM |
| *Research Policy* | Management | PC |
|  | Planning & development | UQ |
| *Science, Technology and Human Values* | Social issues | WM |
| *Science & Public Policy* | Management | PC |
|  | Planning & development | UQ |
|  | Public administration | VM |
| *Minerva* | Education & educationals research | HA |
|  | History & philosophy of science | MQ |
|  | Social science, interdisciplinary | WU |
| *Science Communication* | Communication | EU |
| *R&D Management* | Business | DI |
|  | Management | PC |

**Table 1**: WoS categories for eight STS journals.

Since there is only marginal overlap in the attribution of WCs to these journals most relevant for STS, it is impossible to evaluate an STS unit using WCs for the normalization. Rafols *et al.* (2012) made the point that journal rankings on lists tend to push interdisciplinary units to the margins, using the case of "innovation studies" in the evaluation environment of "business & management." This marginalization is further reinforced when there is no common denominator as in the list of Table 1. Evaluees may be increasingly sensitive to evaluation systems that thus provide institutional incentives to return to disciplinary perspectives (Dahler-Larsen, 2012).

**Conclusion**

The Web-of-Science categories (WCs) have been used increasingly for the selection of reference sets of journals in bibliometric evaluations. One can seriously doubt whether the journals themselves are sufficiently disciplinarily-oriented to be used for the normalization (Boyack & Klavans, 2011). The confusion, however, is potentially aggravated by collapsing sets of journals on grounds which remain otherwise unspecified. We showed the possible problems by providing two examples of the journal sets with which we are most familiar: one discipline that is attributed a WC ("information science & library science") and one specialty (STS) that is not attributed a WC. In both cases, normalizations using these categories might seriously harm the quality of the evaluation. There is no reason to see these fields as exceptions.

What are the alternatives? One would expect the classification systems of Scopus and the alternative offered by Glänzel & Schubert (2003) to do a better job for the simple reason that these classifications of journals were developed in the early 2000s with the explicit aim of bibliometric analysis, whereas the WCs have developed incrementally since the early times of the *Science Citation Index* in the 1960s and 1970s for the purpose of facilitating information retrieval. The WCs have been adapted piecemeal ever since; for example, the number of WCs has grown in JCR from 204 in 1994 to 226 currently,[7] and a number of journals have been reclassified.

The use of (sets of) journals is not the only way to generate reference sets for the evaluation of institutional units or individuals. Alternatively, one can explore the use of professionally developed index terms for the delineation, such as the Medical Subject Headings (MeSH) in Medline/PubMed or *Chemical Abstracts* (e.g., Bornmann *et al*., 2009; Rotolo & Leydesdorff, in press). Another option is to consider the citing papers, that is, the audience of a given paper, as a reference set. This idea is already entertained when counting citations fractionally (Leydesdorff & Bornmann, 2011; Zitt & Small, 2008). One problem, however, may be the "double citation window" thus generated, since one would have to wait until the citing papers are sufficiently cited.

---

[7] Twenty-three more categories are specifically included in the Arts & Humanities Citation Index (A&HCI; Leydesdorff, Hammarfelt, and Salah, 2011).

A more radical approach involves clustering the citations at the level of the papers in the database such as pursued by CWTS (Waltman & Van Eck, 2012), and in the meantime applied in the Leiden Rankings. The latter, for example, are based on normalization against 828 "fields," that is, algorithmically generated clusters of citation relations. Because the fields are algorithmic artifacts, they cannot easily be named (as different from numbered), and therefore cannot be validated. Furthermore, a paper has to be cited or contain references in order to be classified, since the approach is based on direct citation relations; the journal names are used as a second index key to attribute the non-cited papers to the most resembling clusters. Note that Rafols and Leydesdorff (2009) found that algorithmically generated classifications of journals have characteristics very different from content-based classifications (as predicted by Garfield (1955). The Leiden system is not only difficult to validate, it also cannot be accessed or replicated from outside its context of use (cf. Ruiz-Castillo & Waltman, in preparation).

Perhaps, one could also normalize without the specification of reference sets; for example, on the basis of "universal" properties of the distributions (e.g., Radicchi *et al.*, 2008). In summary, opening the metaphor of "best practices" to the challenge of "best possible practices" may provide us with new research perspectives (e.g., Butler, 2010; Colliander, in press). Kostoff and Martinez' (2005) rhetorical question of whether citation normalization is realistic seems to drive a research program in evaluative bibliometrics. However, one may wish to be cautious in suggesting valid normalizations in professional practices until the problems of how to define reference sets are further solved.

**Acknowledgement**
We thank Staša Milojević and Ludo Waltman for comments on a previous version of this paper.

**References**
Bensman, S. J. (2001). Bradford's Law and fuzzy sets: Statistical implications for library analyses. *IFLA Journal, 27*, 238-246.
Bensman, S. J. (2007). Garfield and the impact factor. *Annual Review of Information Science and Technology, 41*(1), 93-155.
Bensman, S. J., & Leydesdorff, L. (2009). Definition and Identification of Journals as Bibliographic and Subject Entities: Librarianship vs. ISI Journal Citation Reports (JCR) Methods and their Effect on Citation Measures. *Journal of the American Society for Information Science and Technology, 60*(6), 1097-1117.
Blondel, V. D., Guillaume, J. L., Lambiotte, R., & Lefebvre, E. (2008). Fast unfolding of communities in large networks. *Journal of Statistical Mechanics: Theory and Experiment, 8*(10), 10008.
Bordons, M., Morillo, F., & Gómez, I. (2004). Analysis of cross-disciplinary research through bibliometric tools. In H. F. Moed, W. Glänzel & U. Schmoch (Eds.), *Handbook of quantitative science and technology research* (pp. 437-456). Dordrecht: Kluwer.
Bornmann, L., & Marx, W. (2014). How to evaluate individual researchers working in the natural and life sciences meaningfully? A proposal of methods based on percentiles of citations. *Scientometrics, 98*(1), 487-509.
Bornmann, L., Marx, W., Schier, H., Rahm, E., Thor, A., & Daniel, H. D. (2009). Convergent validity of bibliometric Google Scholar data in the field of chemistry--Citation counts for papers that were accepted by Angewandte Chemie International Edition or rejected but published elsewhere, using Google Scholar, Science Citation Index, Scopus, and Chemical Abstracts. *Journal of Informetrics, 3*(1), 27-35.

Bornmann, L., Wagner, C., & Leydesdorff, L. (2014; in press). BRICS countries and scientific excellence: A bibliometric analysis of most frequently-cited papers. *Journal of the Association for Information Science and Technology*.

Boyack, K. W., & Klavans, R. (2011). Multiple Dimensions of Journal Specifity: Why journals can't be assigned to disciplines. In E. Noyons, P. Ngulube & J. Leta (Eds.), *The 13th Conference of the International Society for Scientometrics and Informetrics* (Vol. I, pp. 123-133). Durban, South Africa: ISSI, Leiden University and the University of Zululand.

Braun, T., Glänzel, W., Maczelka, H., & Schubert, A. (1994). World science in the eighties. National performances in publication output and citation impact, 1985–1989versus 1980–1984 part I. All science fields combined, physics, and chemistry. *Scientometrics, 29*(3), 299-334.

Butler, L. (2010). *An alternative to WoS subject categories: redefining journal sets for closer alignment to a national classification scheme.* Book of Abstracts of the Eleventh International Conference on Science and Technology Indicators (pp. 62-64), Leiden, September 9-11.

Colliander, C. (2014). A novel approach to citation normalization: A similarity-based method for creating reference sets. *Journal of the Association for Information Science and Technology*.

Costas, R., van Leeuwen, T. N., & Bordons, M. (2010). A bibliometric classificatory approach for the study and assessment of research performance at the individual level: The effects of age on productivity and impact. *Journal of the American Society for Information Science and Technology, 61*(8), 1564-1581.

Garfield, E. (1955). Citation Indexes for Science: A New Dimension in Documentation through Association of Ideas. *Science, 122*(3159), 108-111.

Garfield, E. (1971). The mystery of the transposed journal lists—wherein Bradford's Law of Scattering is generalized according to Garfield's Law of Concentration. *Current Contents, 3*(33), 5–6.

Garfield, E. (1972). Citation Analysis as a Tool in Journal Evaluation. *Science 178*(Number 4060), 471-479.

Gingras, Y., & Larivière, V. (2011). There are neither "king" nor "crown" in scientometrics: Comments on a supposed "alternative" method of normalization. *Journal of Informetrics, 5*(1), 226-227.

Glänzel, W. (2010). On reliability and robustness of scientometrics indicators based on stochastic models. An evidence-based opinion paper. *Journal of Informetrics, 4*(3), 313-319.

Glänzel, W., & Schubert, A. (2003). A new classification scheme of science fields and subfields designed for scientometric evaluation purposes. *Scientometrics, 56*(3), 357-367.

Haustein, S. (2012). *Multidimensional journal evaluation: analyzing scientific periodicals beyond the impact factor*. Berlin/Boston: Walter de Gruyter.

Kamada, T., & Kawai, S. (1989). An algorithm for drawing general undirected graphs. *Information Processing Letters, 31*(1), 7-15.

Katz, J. S., & Hicks, D. (1995). *The classification of interdisciplinary journals: a new approach.* Paper presented at the Proceedings of the Fifth International Conference of the International Society for Scientometrics and Informetrics, Riverforest IL, June 7-10.

Kostoff, R. N., & Martinez, W. L. (2005). Is citation normalization realistic? *Journal of Information Science, 31*(1), 57-61.

Leydesdorff, L. (2006). Can Scientific Journals be Classified in Terms of Aggregated Journal-Journal Citation Relations using the Journal Citation Reports? *Journal of the American Society for Information Science & Technology, 57*(5), 601-613.

Leydesdorff, L. (2007). Mapping Interdisciplinarity at the Interfaces between the *Science Citation Index* and the *Social Science Citation Index*. *Scientometrics, 71*(3), 391-405.

Leydesdorff, L. (2008). *Caveats* for the Use of Citation Indicators in Research and Journal Evaluation. *Journal of the American Society for Information Science and Technology, 59*(2), 278-287.

Leydesdorff, L., & Bornmann, L. (2011). How fractional counting affects the Impact Factor: Normalization in terms of differences in citation potentials among fields of science. *Journal of the American Society for Information Science and Technology, 62*(2), 217-229.

Leydesdorff, L., & Opthof, T. (2013). Citation Analysis using the Medline Database at the Web of Knowledge: Searching "Times Cited" with Medical Subject Headings (MeSH). *Journal of the American Society for Information Science and Technology, 64*(5), 1076-1080.

Leydesdorff, L., Carley, S., & Rafols, I. (2013). Global Maps of Science based on the new Web-of-Science Categories *Scientometrics, 94*(2), 589-593. doi: 10.1007/s11192-012-0783-8

Leydesdorff, L., Hammarfelt, B., & Salah, A. (2011). The structure of the Arts & Humanities Citation Index: A mapping on the basis of aggregated citations among 1,157 journals. *Journal of the American Society for Information Science and Technology, 62*(12), 2414-2426.

Leydesdorff, L., Wagner, C. S., & Bornmann, L. (2014). The European Union, China, and the United States in the top-1% and top-10% layers of most-frequently cited publications: Competition and collaborations. *Journal of Informetrics, 8*(3), 606-617.

Martin, B. R., Nightingale, P., & Yegros-Yegros, A. (2012). Science and technology studies: Exploring the knowledge base. *Research Policy, 41*(7), 1182-1204.

Martin, B., & Irvine, J. (1983). Assessing Basic Research: Some Partial Indicators of Scientific Progress in Radio Astronomy. *Research Policy, 12*, 61-90.

Martyn, J., & Gilchrist, A. (1968). *An Evaluation of British Scientific Journals*. London: Aslib.

Milojević, S., & Leydesdorff, L. (2013). Information Metrics (iMetrics): A Research Specialty with a Socio-Cognitive Identity? *Scientometrics, 95*(1), 141-157.

Milojević, S., Sugimoto, C. R., Larivière, V., Thelwall, M., & Ding, Y. (2014). The role of handbooks in knowledge creation and diffusion: A case of science and technology studies. *Journal of Informetrics, 8*(3), 693-709.

Moed, H., & Van Leeuwen, T. (1995). Improving the Accuracy of the Institute for Scientific Information's Journal Impact Factors. *Journal of the American Society for Information Science, 46*(6), 461-467.

Morillo, F., Bordons, M., & Gómez, I. (2001). An approach to interdisciplinarity through bibliometric indicators. *Scientometrics, 51*(1), 203-222.

National Science Board. (2014). *Science and Engineering Indicators*. Washington, DC: NSF; http://www.nsf.gov/statistics/seind14/.

Neuhaus, C., & Daniel, H.-D. (2009). A new reference standard for citation analysis in chemistry and related fields based on the sections of Chemical Abstracts. *Scientometrics, 78*(2), 219-229.

Nicolaisen, J. (2007). Citation analysis. *Annual review of information science and technology, 41*(1), 609-641.

Opthof, T., & Leydesdorff, L. (2010). *Caveats* for the journal and field normalizations in the CWTS ("Leiden") evaluations of research performance. *Journal of Informetrics, 4*(3), 423-430.

Price, D. d. S. (1970). Citation Measures of Hard Science, Soft Science, Technology, and Nonscience. In C. E. Nelson & D. K. Pollock (Eds.), *Communication among Scientists and Engineers* (pp. 3-22). Lexington, MA: Heath.

Pudovkin, A. I., & Garfield, E. (2002). Algorithmic procedure for finding semantically related journals. *Journal of the American Society for Information Science and Technology, 53*(13), 1113-1119.

Radicchi, F., Fortunato, S., & Castellano, C. (2008). Universality of citation distributions: Toward an objective measure of scientific impact. *Proceedings of the National Academy of Sciences, 105*(45), 17268-17272.

Rafols, I., & Leydesdorff, L. (2009). Content-based and Algorithmic Classifications of Journals: Perspectives on the Dynamics of Scientific Communication and Indexer Effects. *Journal of the American Society for Information Science and Technology, 60*(9), 1823-1835.

Rafols, I., Leydesdorff, L., O'Hare, A., Nightingale, P., & Stirling, A. (2012). How journal rankings can suppress interdisciplinary research: A comparison between innovation studies and business & management. *Research Policy, 41*(7), 1262-1282.

Rehn, C., Gornitzki, C., Larsson, A., & Wadskog, D. (2014). *Bibliometric Handbook for Karolinska Institutet*. Stockholm: Karolinska Institute.

Rotolo, D., & Leydesdorff, L. (in press). Matching MEDLINE/PubMed Data with Web of Science (WoS): A Routine in R-language. *Journal of the Association for Information Science and Technology*.

Ruiz-Castillo, J., & Waltman, L. (2014, in preparation). Field-normalized citation impact indicators using algorithmically constructed classification systems of science.

Schubert, A., & Braun, T. (1986). Relative indicators and relational charts for comparative assessment of publication output and citation impact. *Scientometrics, 9*(5), 281-291.

Schubert, A., Glänzel, W., & Braun, T. (1989). Scientometric datafiles. A comprehensive set of indicators on 2649 journals and 96 countries in all major science fields and subfields 1981–1985. *Scientometrics, 16*(1), 3-478.

van Leeuwen, T. N., & Medina, C. C. (2012). Redefining the field of economics: Improving field normalization for the application of bibliometric techniques in the field of economics. *Research Evaluation, 21*(1), 61-70.

Waltman, L., & Van Eck, N. J. (2012). A new methodology for constructing a publication-level classification system of science. *Journal of the American Society for Information Science and Technology, 63*(12), 2378-2392.

Waltman, L., Calero-Medina, C., Kosten, J., Noyons, E., Tijssen, R. J. W., van Eck, N. J., . . . Wouters, P. (2012). The Leiden Ranking 2011/2012: Data collection, indicators, and interpretation. *Journal of the American Society for Information Science and Technology, 63*(12), 2419-2432.

Waltman, L., Van Eck, N. J., Van Leeuwen, T. N., Visser, M. S., & Van Raan, A. F. J. (2011). Towards a New Crown Indicator: Some Theoretical Considerations. *Journal of Informetrics, 5*(1), 37-47.

Waltman, L., Yan, E., & van Eck, N. J. (2011). A recursive field-normalized bibliometric performance indicator: An application to the field of library and information science. *Scientometrics, 89*(1), 301-314.

Wouters, P. (1998). The signs of science. *Scientometrics, 41*(1), 225-241.

Zitt, M., & Small, H. (2008). Modifying the journal impact factor by fractional citation weighting: The audience factor. *Journal of the American Society for Information Science and Technology, 59*(11), 1856-1860.